\documentstyle[twoside,fleqn,espcrc2,fps]{article}
\newcommand{\bibi}{\bibitem}

\newcommand{\gm}{\gamma}

\newcommand{\et}{\eta}

\newcommand{\kp}{\kappa}

\newcommand{\ps}{\psi}

\newcommand{\nnn}{\nonumber \\}

\newcommand{\Om}{\Omega}

\newcommand{\psb}{\overline{\ps}}

\newcommand{\chb}{\overline{\chi}}

\newcommand{\hmu}{\hat{\mu}}

\newcommand{\eighth}{\mbox{{\small $\frac{1}{8}$}} }

\newcommand{\IM}{\mbox{Im\,}}
\newcommand{\RE}{\mbox{Re\,}}

\newcommand{\dg}{\dagger}
\newcommand{\pr}{\prime}
\newcommand{\ra}{\rightarrow}

\newcommand{\be}{\begin{equation}}
\newcommand{\ee}{\end{equation}}
\newcommand{\bea}{\begin{eqnarray}}
\newcommand{\eea}{\end{eqnarray}}
\newcommand{\eq}{\ref}
\newcommand{\beq}{\begin{equation}}
\newcommand{\eeq}{\end{equation}}
\newcommand{\cc}{\cite}
\newcommand{\lb}{\label}
\newcommand{\lsim}{\stackrel{<}{\sim}} 

\def \3{\ss}

\title{Chiral gauge theories on the lattice and restoration of
gauge symmetry}

\author{
Wolfgang Bock\address{
University of California, San Diego, Department of Physics,\\
9500 Gilman Drive 0319, La Jolla, CA 92093-0319, USA}%
\thanks{Speaker at the conference},
Jan Smit\address{
Institute of Theoretical Physics, University of Amsterdam,\\
Valckenierstraat 65, 1018 XE Amsterdam, The Netherlands}
and Jeroen C. Vink$^{{\rm a}}$
}

\begin{document}
\begin{abstract}
We investigate a proposal for the construction
of models with chiral fermions on the lattice using
staggered fermions. In this approach the gauge invariance is
broken by the coupling of the staggered fermions to the gauge fields.
We aim at a dynamical restoration of the gauge invariance in the full
quantum model. If the gauge symmetry breaking (SB) is not too severe,
this procedure could lead in the continuum limit to the desired gauge
invariant chiral gauge theory.
\end{abstract}
%
\maketitle
\section{Introduction}
The non-perturbative formulation of chiral gauge theories
is still an unsolved problem. For a review on recent attempts
in this direction see ref.~\cc{REV}.
In this contribution we will focus on the staggered fermion approach.
The basic idea here is to use the species doublers as
physical degrees of freedom, rather than to try to decouple them.
Staggered fermions are represented on the lattice by a one-component
complex field $\chi_x$, which
describes four Dirac flavors in the continuum limit.
The Dirac and flavor components
of these four staggered flavors do not appear in an explicit form
since they are spread out over the lattice and it is not clear
a priori how to construct actions with arbitrary spin-flavor coupling.
It has been shown in ref.~\cc{Jan} that this can be  achieved
by first rewriting the naive action in terms of the
the $4 \times 4$ matrix fields $\Psi^{df}$
with $d$ and $f$ acting as Dirac and flavor indices and then using
the relation
\be
\Psi_x^{df} =\eighth \sum_{b}  [\gm^{x+b}]^{df}
       \chi_{x+b} \lb{PSI-CHI}
\ee
to express the action in terms of the independent $\chi$-fields.
The sum in (\eq{PSI-CHI}) is over the 16 corners of a lattice
hypercube, $b_{\mu}=0,1$,  and
$\gm^x \equiv \gm_1^{x_1} \gm_2^{x_2} \gm_3^{x_3}\gm_4^{x_4}$.
It is evident from (\eq{PSI-CHI})
that the $16$ spin-flavor components of $\Psi_x$ are not independent.
One can show however that the components of the Fourier
transform  $\Psi(p)$ are independent in the restricted
momentum interval $-\pi/2 < p_{\mu} \leq  +\pi/2$.
The target model which we will consider in this paper
is a four flavor axial-vector
model with all axial charges $q$ equal to $+1$.
For the  action in terms of the
$\chi$-fields we find     the following expression
\bea
  S \!\!\!&=&\!\!\! - \frac{1}{2} \sum_{x\mu}\left[ c_{\mu x}
 \frac{1}{16}\sum_b \et_{\mu x+b} \right. \nnn
 \!\!\!&&\!\!\!\!   \times (\chb_{x+b} \chi_{x+b+\hmu}
 - \chb_{x+b+\hmu} \chi_{x+b})  \nnn
 \!\!\!  &+&\!\!\!   i s_{\mu x} \frac{1}{16}\sum_{b+c=n}
 \et_{5 x+b} (\et_{\mu x+c}
      \chb_{x+b} \chi_{x+c+\hmu}  \nonumber \\
\!\!\!&&\!\!\!-  \left. \et_{\mu x+b} \chb_{x+b+\hmu} \chi_{x+c})
     \phantom{ \frac{1}{2}} \!\!\! \right] \;,  \label{NISF}
\eea
where $c_{\mu x}=\RE U_{\mu x}$,
$s_{\mu x}=\IM U_{\mu x}$,   $n= ( 1,$ $  1, 1, 1 )$.
The sign factors
$\eta_{\mu x}=(-1)^{x_1+ \ldots + x_{\mu-1}}$
and $\et_{5 x}=-\et_{4 x} \;\et_{3 x+\hat{4}}
\;\et_{2 x+\hat{3}+\hat{4}}\;\et_{1 x+\hat{2}+\hat{3}+\hat{4}} $
represent here the $\gm_{\mu}$ and $\gm_5$. It can be shown easily
that (\eq{NISF}) reduces in the classical continuum limit to the
target model. The expression (\eq{NISF})
however lacks gauge invariance,
since gauge transformations on the $U$-fields cannot
be carried through to the $\chi$-fields.
A perturbative calculation in two dimensions
has shown that the model performs well
for smooth external gauge fields \cc{SMOOTH}.
The important issue however is whether this remains true also when
taking into account the full quantum fluctuations. Based on experience
with other non-gauge invariant lattice models  we shall aim here for
a dynamical symmetry restoration (SR), which may occur
if the SB effects due the high momentum modes
are not too severe.
\section{Restoration of gauge symmetry}
Let's start first from a generic non-gauge invariant lattice action,
$S(U)$. It has been shown in ref.~\cc{Jan}
that the partition function can be written in the following form
$Z = \int D U e^{  S(U_{\mu x}) }
   =  \int D U D V e^{ S(V_x^{\dg} U_{\mu x} V_{x+\hmu})}$,
where the $V_x$ is a radially frozen Higgs field.
The new form of the action $S(V_x^{\dg} U_{\mu x} V_{x+\hmu})$
is  trivially invariant under the local gauge transformations
$U_{\mu x} \ra \Om_x U_{\mu x} \Om^{\dagger}_{x+\hmu}$,
$V_x \ra \Om_x     V_x$.
The important question of symmetry restoration
is whether the resulting model
still can describe the physics of the underlying gauge invariant model which
in our case is the chirally invariant axial-vector vector model
in the continuum.
%
%
\begin{figure}[t]
 \centerline{
 \fpsysize=8.0cm
 \fpsbox{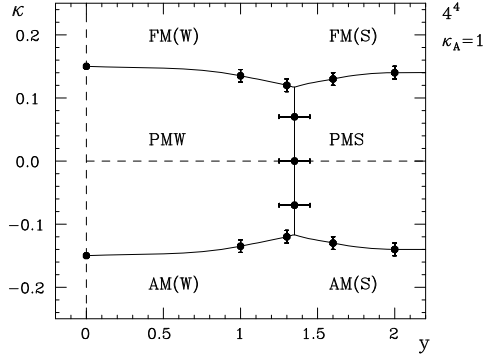}
 }
\vspace*{-1.5cm}
\caption{ \noindent {\em Phase diagram of the ISF model.
}}
\label{FIG1}
\end{figure}
Before we try to find an answer to  this  question for the staggered model,
let us first give     an example  of a non-chiral lattice model
where gauge symmetry gets restored.
The model is given by the
action
\bea
S \!\!\!&=&\!\!\! -\frac{1}{2}
\sum_{x \mu} \left\{ \psb_x^\pr \gm_{\mu}
( U_{\mu x} P_L + U_{\mu x}^*  P_R )
\psi_{x+\hmu}^\pr -  \right. \nnn
\!\!\!&&\!\!\!\!\!\!\!\!\!\!\!\!\!\!\!\!\!\!\!
\left. \psb_{x+\hmu}^\pr \gm_{\mu}
(U_{\mu x}^* P_L+U_{\mu x} P_R)
\psi_x^\pr \right\}
- y \sum_x \psb_x^\pr \psi_x^\pr \;, \lb{NAIV}
\eea
which for $y=0$ reduces in the classical
continuum limit to a
gauge invariant $8(q=+1)+8(q=-1)$ axial-vector
model. This model we shall regard now  as
our target model. The mass term in (\eq{NAIV}) breaks
gauge invariance. The action,
which results after integrating over the
gauge fields in the partition function
is given by (\eq{NAIV}) with
$U_{\mu x} \ra  V_x^* U_{\mu x} V_{x+\hmu}$.
The $\psi^\pr$-fields  in this action are screened form
the gauge fields by the $V$-fields and
are therefore {\em neutral} with respect to
the U(1) gauge transformations. To see
how the symmetry gets restored it is useful
to express the action in terms of a {\em charged} fermion field $\psi$
which is  related to the $\psi^\pr$-fields by a gauge transformation,
$ \psi^\pr_x  = (V_x^* P_L+V_x P_R) \psi_x $. The $\psi$-action
is identical with (\eq{NAIV}) (with $\psi^\pr \ra \psi$), except
that the bare mass term turns into a Yukawa-term
$- y \sum_x \psb_x (V_x^2 P_L+ {V_x^*}^2 P_R) \psi_x $ with
a charge two Higgs field $V_x^2$. We have studied this  Yukawa-model
in the global symmetry limit, $U_{\mu x}=1$, and with
the term $2\kp \sum_{x \mu} \RE ( V_x^*  V_{x+\hmu}) $
added to the action. The model we have
studied numerically is a
$4(q=+1)+4(q=-1)$ model which results after
a   mirror fermion doubling needed for the Hybrid Monte Carlo
algorithm  and a standard fermion
number reduction.
This model we shall call
the invariant staggered fermion (ISF)
model, to emphasize that the $\psi$-and $\psi^\pr$-forms
are related by a gauge transformation.
The $\kp$-$y$ phase diagram of the ISF model is
displayed in fig.~\ref{FIG1}.
Besides the ferromagnetic (FM) and antiferromagnetic
(AM) phases
there are two different symmetric phases, PMW and PMS.
Previous  investigations  have shown that the physics in the
PMW (PMS) phase is described by the action in terms of the
$\psi$ ($\psi^\pr$)-fields  and  that
the fermions are massless (massive) in the PMW (PMS) phase.
The gauge symmetry gets restored if the coefficient
$y$ of the SB term is $ < y_c(\kp) \approx 1.4$:
Anywhere in the PMW phase, away from the
phase boundaries, the scalar particles
decouple and even though $y>0$, the spectrum contains only free
massless fermions.
The effective action in this region (with gauge fields switched
on) is given by (\eq{NAIV}) with the SB  mass
term dropped.
In contrast for $y > y_c(\kp)$ fermions are massive and do
not couple to the gauge fields. In this region
we do not recover the physics of the target model.
%
%
%
\begin{figure}[t]
\centerline{
\fpsysize=8.0cm
\fpsbox{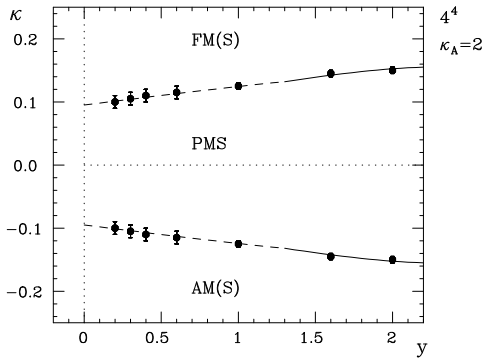}
}
\vspace*{-1.5cm}
\caption{ \noindent {\em Phase diagram of the NISF model.
}}
\label{FIG2}
\end{figure}
\section{SR in the staggered model?}
The gauge invariant version of the staggered fermion
model with the Higgs fields included is
given by (\eq{NISF}) with
$c_{\mu x} \ra \RE (V_x^* U_{\mu x} V_{x+\hmu})$ and
$s_{\mu x} \ra \IM (V_x^* U_{\mu x} V_{x+\hmu})$. It is clear that
the $\chi$-fields are neutral
with respect to the U(1) transformations which
allows us also to add a term $-y \sum_x \chb_x \chi_x$ to the
action. In contrast to the
ISF model
we cannot relate the
$\chi$-fields by a gauge transformation to a charged fermion
field. The crucial question
however is whether the Higgs fields are sufficiently
smooth such that this   transformation
could effectively take place, leading to a PMW
phase at small $y$ with a charged massless fermion (then gauge
symmetry would be restored dynamically).
The gauge symmetry is broken
now due to the high momentum modes  in the $V$-fields and
the strength of the gauge SB cannot be easily controlled by a
parameter in the action (like
$y$ in (\eq{NAIV})). We hope however that the SB effects are not very
severe and a PMW phase, where the gauge symmetry gets restored,
emerges at small $y$.
For technical reasons the numerical simulations have been performed
in a vector-like
$4(q=+1)+4(q=-1)$ model which is obtained from (\eq{NISF}) after
a mirror doubling and  which we shall call the
non-invariant staggered fermion (NISF) model.
The phase diagram of this model contains indeed at small $y$ a phase with
very bizarre properties, not at all characteristic for a PMW phase \cc{ST4}.
Its existence could be  ascribed
to the hypercubic average in the $\gm_{\mu}$-term in (\eq{NISF}). To make
the model more similar to the ISF model we
have dropped the $\sum_b$ in this term
which does not change the classical continuum limit \cc{ST4}.
The $\kp$-$y$ phase diagram
of this modified model is shown in fig.~2. It can be seen that the PMS phase
now extends down to $y=0$. Also on a larger lattice, which can accommodate a
larger number of small momentum modes we found no indication for the emergence
of a PMW phase.  Another possibility to enhance the low
momentum modes of the scalar field,
is to increase the value of $\kp$ towards the  FM-PM
phase transition where the scalar field correlation length may become large
enough that a PMW phase opens up.
The FM(S)-PMS phase transition for $y \lsim 1.3$ is however of first order
and the scalar field correlation length cannot exceed a certain bound.
Our runs at $\kp \approx \kp_c^{FM-PM}$ show indeed no qualitative
difference from the results at other $\kp < \kp_c^{FM-PM}$.
This negative result means that more sophisticated strategies have
to be developed to make the SR working.
As an alternative one can  start also from a
gauge fixed continuum model and regularize it using the lattice and
staggered fermions. The model then becomes very
similar to the Rome proposal, which uses Wilson fermions \cc{ROME}.

The numerical calculations were performed on the CRAY Y-MP4/464
at SARA, Amsterdam. This research was supported by the ``Stichting voor
Fun\-da\-men\-teel On\-der\-zoek der Materie (FOM)'',
by the ``Stichting Nationale Computer Faciliteiten (NCF)'' and by
the DOE under contract DE-FG03-90ER40546.

\end{document}